\documentclass[12pt,preprint,useAMS,usenatbib]{emulateapj}
\usepackage{graphicx}
\usepackage{amsmath}
\usepackage{color}

\newcommand{\Msun}{\,\rm M_\odot}
\newcommand{\Zsun}{\rm Z_\odot}
\newcommand{\Zsn}{\rm Z_{SN}}

\newcommand{\f}{\frac} 
\newcommand{\Rvir}{R_{\rm vir}}

\newcommand{\HI}{H\textsc{i}}
\newcommand{\HH}{H$_2$}
\newcommand{\fHH}{f_{{\rm H}_2}}
\newcommand{\SigmaSFR}{\Sigma_{\rm SFR}}
\newcommand{\SigmaGas}{\Sigma_{\rm gas}}

\newcommand{\rhoGas}{\rho_{\rm gas}}
\newcommand{\fstar}{\rm f_\star} 
\newcommand{\fdark}{\rm f_{\rm dark}} 
\newcommand{\Mhalo}{\rm M_h}

\newcommand{\Mstar}{\rm M_\star}
\newcommand{\MHI}{\rm M_{HI}}

\newcommand{\OmegaHI}{\rm \Omega_{HI}}
\newcommand{\tdep}{\rm t_{dep}} 
\newcommand{\tdepHI}{\rm t_{dep,HI}} 
\newcommand{\Zfloor}{\rm Z_{floor}}
\newcommand{\LUV}{\rm L_{UV}}

\begin{document}

\title{Dwarf Galaxy Formation with \HH-Regulated Star Formation: II. Gas-Rich Dark Galaxies at Redshift 2.5}
\author{Michael Kuhlen$^{1,2}$, Piero~Madau$^2$, and Mark Krumholz$^2$}

\altaffiltext{1}{Theoretical Astrophysics Center, University of California Berkeley, Hearst Field Annex, Berkeley, CA 94720.}
\altaffiltext{2}{Department of Astronomy and Astrophysics, University of California, 1156 High Street, Santa Cruz, CA 95064.}

\begin{abstract}
We present a cosmological hydrodynamic simulation of the formation of dwarf galaxies at redshifts $z \gtrsim 2.5$ using a physically-motivated model for \HH-regulated star formation. Our simulation, performed using the Enzo code and reaching a peak resolution of 109 proper parsecs at $z=2.5$, extends the results of \citet{kuhlen_dwarf_2012} to significantly lower redshifts. We show that a star formation prescription  regulated by the local \HH\ abundance leads to the suppression of star formation in dwarf galaxy halos with $\Mhalo \lesssim 10^{10}\,\Msun$ and to a large population of gas-rich ``dark galaxies" at $z=2.5$ with low star formation efficiencies and gas depletion timescales $>20$ Gyr. The fraction of dark galaxies is 60\% at $\Mhalo\simeq 10^{10}\,\Msun$ and increases rapidly with decreasing halo mass. Dark galaxies form late and their gaseous disks never reach the surface densities, $\gtrsim 5700 \Msun \, {\rm pc}^{-2}\, (Z/10^{-3}\Zsun)^{\!-0.88}$, that are required to build a substantial molecular fraction. Despite this large population of dark galaxies, we show that our \HH-regulated simulation is consistent with both the observed luminosity function of galaxies and the cosmological mass density of neutral gas at $z\gtrsim 2.5$. Moreover, our results provide a theoretical explanation for the recent detection in fluorescent Ly$\alpha$ emission of gaseous systems at high redshift with little or no associated star formation. We further propose that \HH-regulation may offer a fresh solution to a number of outstanding ``dwarf galaxy problems" in $\Lambda$CDM. In particular, \HH-regulation leads galaxy formation to become effectively stochastic on mass scales of $\Mhalo\sim 10^{10}\,\Msun$, and thus these massive dwarfs are not ``too big to fail". 
\end{abstract}

\keywords{cosmology: theory -- galaxies: dwarfs -- galaxies: formation -- galaxies: halos -- methods: numerical}

\section{Introduction}\label{intro}

For almost two decades observations of dwarf galaxies have challenged the $\Lambda$CDM paradigm of cosmological structure formation and our understanding of the
mapping from dark matter halos to their baryonic components. Dwarfs -- either in the field or in the halos of larger systems like our own Milky Way -- are much less abundant than the $\Mhalo \lesssim 10^{10}\,\Msun$ dark matter halos that should, in principle, be able to host them. Dark matter-only simulations predict steep (``cuspy") inner density profiles, while the observed rotation curves of dwarf galaxies instead suggest that they have near-constant density cores. Similarly, the most massive subhalos found in dark matter-only simulations of Milky Way-sized systems appear to be too dense to be consistent with existing constraints on the classical Galactic dwarf satellites.

While the discrepancies above may be truly reflective of a fundamental failure of the otherwise remarkably successful $\Lambda$CDM model, numerous astrophysical solutions have been proposed to explain this ``dwarf galaxy problem" and are being actively investigated. To date, they all appear to provide only a partial, often environment-dependent solution to the puzzle. Photo-heating from the cosmic ultraviolet background may suppress gas infall into dwarf galaxy halos in the low-redshift universe \citep{efstathiou_suppressing_1992,thoul_hydrodynamic_1996}, but the importance of photoionization feedback is greatly reduced at high redshift because dwarf galaxy-sized objects can either self-shield against the ionizing background \citep{dijkstra_photoionization_2004} or accrete substantial amount of gas prior to reionization \citep{bullock_reionization_2000,madau_fossil_2008}. Gas removal by ram pressure in massive galaxy halos \citep{mayer_simultaneous_2006} or through hydrodynamical interactions with the cosmic web \citep{benitez-llambay_dwarf_2013}, while able to quench star formation in satellite and isolated dwarfs at late times, must operate in concert with other suppression mechanisms that prevent the early conversion of gas into stars in such systems. Mass loss driven by supernova (SN) feedback can reduce the baryonic content of bright dwarfs \citep{dekel_origin_1986,mori_early_2002,governato_forming_2007} and flatten their central dark matter cusps \citep{read_mass_2005,mashchenko_stellar_2008,governato_bulgeless_2010,pontzen_how_2012,teyssier_cusp-core_2013}, in all but the most star deficient dwarf spheroidals \citep{garrison-kimmel_can_2013,zolotov_baryons_2012,boylan-kolchin_milky_2012,penarrubia_coupling_2012}.

In \citet[Paper I]{kuhlen_dwarf_2012} we followed a different avenue to quench star formation in dwarfs, one based on a new understanding of the chemistry and thermodynamics of the interstellar gas that is actually forming stars. Spatially resolved observations of local galaxies have revealed that star formation correlates much more tightly with the density of molecular gas than total gas density \citep{leroy_star_2008,bigiel_star_2008}, even in regions where molecular gas constitutes only a trace component of the ISM \citep{schruba_molecular_2011}. Even though the primary cooling agents are  lines of CO or C~\textsc{ii} (depending on the chemical state of the carbon), molecular hydrogen (\HH) is expected to be good tracer of star formation \citep{krumholz_which_2011, glover_is_2012, glover_star_2012}. This is because star formation occurs only where the gas is able to reach very low temperatures, which is possible only in regions where the extinction is high enough to block out the background interstellar radiation field that is responsible both for heating the gas and for dissociating H$_2$ molecules.

In Paper I we showed that \HH-regulated star formation leads to the suppression of star formation in dwarf galaxy halos at $z \gtrsim 4$, and discussed how such a novel quenching mechanism -- one that modifies the efficiency of star formation of cold gas directly, rather than indirectly reducing the cold gas content with supernova feedback -- may contribute to alleviate some of the issues faced by theoretical galaxy formation models. Similar models have been investigated numerically by \citet{gnedin_modeling_2009}, \citet{gnedin_kennicutt-schmidt_2010, gnedin_environmental_2011} and \citet{christensen_implementing_2012}, and semi-analytically by \citet{fu_atomic--molecular_2010} and \citet{krumholz_metallicity-dependent_2012}. Our work in Paper I and here differs from the previous numerical simulations in that we investigate a cosmologically-representative volume rather than single galaxies, and we are therefore able to discuss the statistical properties of the galaxy population produced by \HH-regulated star formation. At the same time, we are still performing full three-dimensional simulations, rather than relying on semi-analytic model prescriptions. Here, we extend our state-of-the-art simulations to lower redshifts and confirm the existence at these epochs of a large population of very inefficient star formers below $\Mhalo\simeq 10^{10}\,\Msun$. We show that such ``dark galaxies" form late and that their gaseous disks never reach the surface densities required to build a substantial molecular fraction.

\section{Simulation and Molecular Chemistry}

We summarize here the main features of our cosmological AMR hydrodynamics simulation (see \citet{kuhlen_dwarf_2012} for a more extensive discussion) with
the Enzo v2.2 code\footnote{http://enzo-project.org/}. The computational domain
covers a (12.5 Mpc)$^3$ box with a root grid of $256^3$ grid
cells. The dark matter density field is resolved with $256^3$
particles of mass $3.1 \times 10^6 \Msun$. Adaptive mesh refinement is
allowed to occur throughout the entire domain for a maximum of 7
levels of refinement, resulting in a maximum spatial resolution of
$\Delta x_7 = 109 \times 3.5/(1+z)$ proper parsec. Mesh refinement is
triggered by a grid cell reaching either a dark matter mass equal to 4
times the mean root grid cell dark matter mass, or a baryonic mass
equal to $8 \times 2^{-0.4 l}$ times the mean root grid cell baryonic
mass, where $l$ is the grid level. The simulation is initialized at $z=99$ 
with cosmological parameters consistent with the WMAP 7-year results
\citep{komatsu_seven-year_2011}. It includes radiative cooling from both primordial and
metal enriched gas, as well as photo-heating from an optically thin,
uniform meta-galactic UV background. An artificial pressure support is applied to cells that have reached the
maximum refinement level in order to stabilize these cells against artificial fragmentation. 

We implement an \HH-regulated star formation prescription as follows. The local star formation rate (hereafter SFR) 
in a grid cell is proportional to the molecular hydrogen density by the free-fall time, $t_{\rm ff}$, 
determined from the total gas density, 
\begin{equation}
\dot{\rho}_{\rm SF} = \epsilon \, \fHH \f{ \rhoGas }{ t_{\rm ff} },
\end{equation}
where $\fHH = \rho_{{\rm H}_2}/\rho_{\rm gas}$ is the \HH\ fraction, 
computed as described below, and the star formation efficiency 
$\epsilon = 0.01$, the value favored by observations \citep{krumholz_slow_2007, krumholz_universal_2012}.
No density threshold for 
star formation is applied. Stars form only once per root grid time step and only
in cells at the highest refinement level (here $l=7$), 
but the star particle mass is proportional to the root grid time step
$\Delta t_0$ \citep[see][]{kravtsov_origin_2003}, i.e.
\begin{equation}
m_* = \epsilon \, \fHH \rhoGas \, (\Delta x_7)^3 \, \f{\Delta t_0}{t_{\rm ff}}.
\end{equation}
We enforce a minimum stellar mass of $m_{\rm min} = 10^4 \Msun$, since even
$\Delta t_0$ can occasionally become very small. Below this mass we
implement a stochastic star formation criterion as follows: if $m_* < m_{\rm min}$, 
we form a particle of mass equal to $m_{\rm min}$ if a
randomly generated number is smaller than $(m_*/m_{\rm min})$.

\begin{figure*}[htp]
\includegraphics[width=\textwidth]{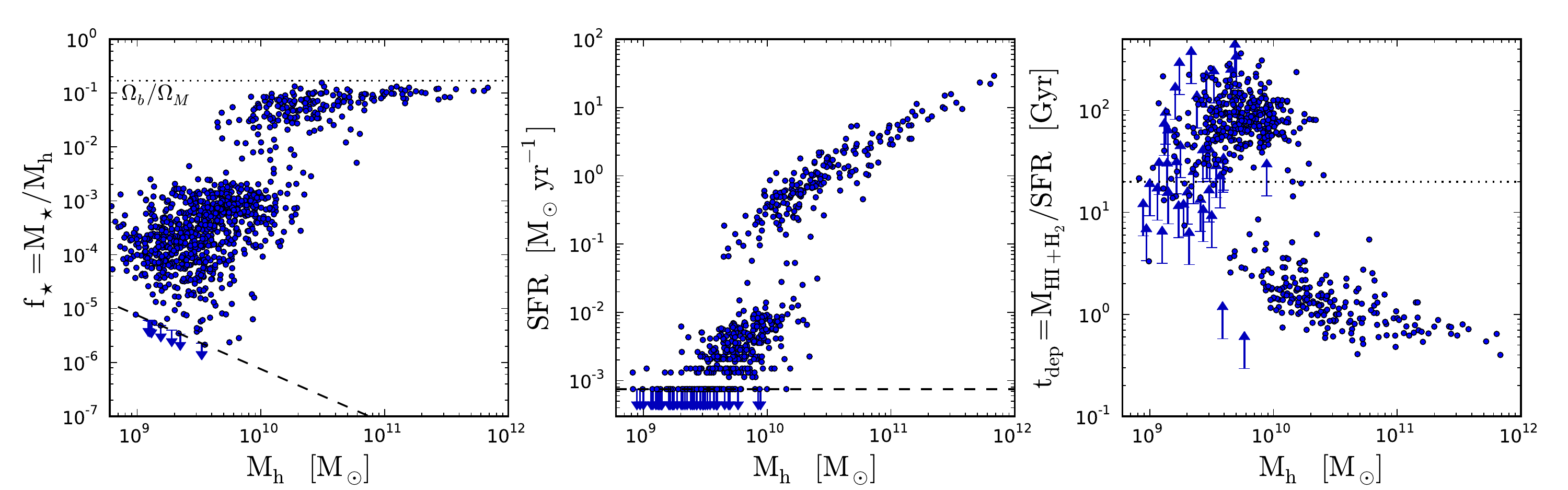}
\caption{Stellar mass fraction $\fstar = M_\star / \Mhalo$ ({\it left panel}), star formation rate ({\it middle panel}), and gas depletion time $\tdep = M_{\rm HI+H_2}/{\rm SFR}$ ({\it right panel}) vs. total halo mass, all at $z=2.5$. In the left panel, the dotted horizontal line indicates an $\fstar$ equal to the cosmic baryon fraction $\Omega_{\rm b}/\Omega_{\rm M}$,  ({\it right panel}) vs. total halo mass. In the left panel, the dotted horizontal line indicates an $\fstar$ equal to the cosmic baryon fraction $\Omega_{\rm b}/\Omega_{\rm M}$, i.e. a 100\% gas to star conversion efficiency. The dashed lines in the left and middle panels indicate the lowest $\fstar$ and SFR that our simulation can resolve, given our minimum stellar particle mass of $(1 - f_{\rm ej}) \, m_{\rm sp} = 7.5 \times 10^3 \Msun$ and SF averaging timescale of 10 Myr. In the right panel the horizontal dotted line marks our definition of a ``dark galaxy'', i.e. one with $\tdep > 20$ Gyr. Completely dark halos ($\rm M_\star=0$, SFR$=0$, or $\tdep=\infty$) are indicated by arrows at the location corresponding to the lowest resolvable $M_\star$ or SFR (for clarity we only plotted 1/10 of all such halos).}
\label{fig:fstar}
\end{figure*}

To obtain the molecular hydrogen mass fraction $\fHH$ in a given grid cell, we use the
two-phase equilibrium model of \citet{krumholz_atomic--molecular_2008}, 
\citet{krumholz_atomic--molecular_2009}, and \citet{mckee_atomic--molecular_2010}, 
hereafter referred to as the KMT model. This model is based on a
radiative transfer calculation of an idealized spherical giant
atomic--molecular complex, subject to a uniform and isotropic
Lyman-Werner (LW) radiation field. The \HH\ abundance is calculated
assuming formation-dissociation balance and a two-phase equilibrium 
between a cold neutral medium (CNM) and a warm neutral medium (WNM) 
\citep{wolfire_neutral_2003}:
\begin{equation}
\fHH \simeq 1 - \f{3}{4} \f{s}{1 + 0.25 s}, \label{eq:KMT09model_begin}
\end{equation}
\begin{equation}
s = \f{\ln(1 + 0.6\chi + 0.01 \chi^2)}{0.6\,\tau_c},
\end{equation}
\begin{equation}
\chi = 2.3 \left( \f{\sigma_{d,-21}}{\mathcal{R}_{-16.5}} \right) \f{1 + 3.1 \, ({\rm Z}/\Zsn)^{0.365}}{\phi_{\rm CNM}}.
\end{equation}
$\tau_c \simeq 0.067 \, (Z/\Zsun) \, (\Sigma_{\rm HI} / \Msun \, {\rm pc}^{-2})$ is the dust optical depth of the cloud, $\sigma_{d,-21}$ is the dust cross-section per H nucleus to 1000
\AA\ radiation, normalized to a value of $10^{-21}$ cm$^{-2}$,
$\mathcal{R}_{-16.5}$ is the rate coefficient for \HH\ formation on
dust grains, normalized to the Milky Way value of $10^{-16.5}$ cm$^3$
s$^{-1}$ \citep{wolfire_chemical_2008}, $\Zsn$ is the gas 
phase metallicity in the solar neighborhood, and $\phi_{\rm CNM}$ is the ratio of the typical CNM density to the minimum density at which a two-phase CNM-WNM equilibrium can be established.  We set $\Zsn = \Zsun$ 
\citep{rodriguez_oxygen_2011}, $\Zsun = 0.0204$, $\phi_{\rm CNM}=3$, and $(\sigma_{d,-21} / \mathcal{R}_{-16.5}) = 1$.
Note that in this model the \HH\ fraction is \textit{independent} of the intensity of the ambient UV radiation field (which is responsible for both heating and \HH\ 
dissociation), since under the assumption of a two-phase CNM-WNM equilibrium the CNM density scales approximately linearly with the UV intensity, and the equilibrium 
\HH-abundance depends on the ratio of the two. \citet{krumholz_comparison_2011} demonstrated that this simple analytical model very accurately reproduces the \HH\ 
abundance determined in full non-equilibrium radiative transfer calculations when $Z/\Zsun \gtrsim 10^{-2}$. At lower metallicities the equilibrium model tends 
to \textit{overpredict} the \HH\ abundance, but at such low metallicities star formation should nevertheless scale with 
the equilibrium \HH\ abundance, not with the non-equilibrium value \citep{krumholz_star_2012}.

While in the KMT model $\fHH$ drops to zero at a metallicity-dependent critical column density of
\begin{eqnarray}
\Sigma_{\rm crit}(Z) & = & \f{\log(1 + 0.6 \chi + 0.01 \chi^2)}{0.0804 \, Z/\Zsun} \, \Msun \, {\rm pc}^{-2} \nonumber \\
              & \approx & 5700 \Msun \, {\rm pc}^{-2} \, \left( \f{Z/\Zsun}{10^{-3}} \right)^{\!-0.88},
\end{eqnarray}
observations indicate that such a hard cutoff is too extreme. In particular, \citet{bigiel_extremely_2010} have demonstrated that the scaling between total gas and SFR surface density continues below the turn-over commonly attributed to the transition from atomic to molecular gas phase ($\lesssim$ few $\Msun$ pc$^{-2}$ at solar metallicity), albeit at a $\sim 50-100$ times lower amplitude. Assuming the molecular gas-SFR scaling relation measured by \citet{schruba_molecular_2011} continues to correspondingly low values of $\SigmaSFR$ ($\lesssim 10^{-4} \Msun \, {\rm yr}^{-1} \, {\rm kpc}^{-2}$), the gas at these extremely low total gas columns has an effective molecular gas fraction of $\fHH \approx 0.01$. In order to capture this empirical low surface density behavior, we impose a a floor of $\fHH = 0.01$ in cells with $T < 10^4$ K \textit{even if they have $\SigmaGas < \Sigma_{\rm crit}(Z)$.}

The resolution of our simulation is not sufficient to resolve the
formation sites of the first generation of stars, the so-called
Population III. In order to capture the metal enrichment resulting
from the supernova explosions of this primordial stellar population,
we instantaneously introduce a metallicity floor of $[\Zfloor] \equiv
\log_{10}(\Zfloor/\Zsun) = -3.0$ at $z=10$, as motivated by recent high
resolution numerical simulations of the transition from Pop III to
Pop II SF \citep{wise_birth_2012}. This ensures the presence of a
minimum amount of metals, which seed subsequent star formation and
further metal enrichment. 

The simulation was run on the NASA supercomputer \textit{Pleiades} and cost $\sim 200,000$ core hours to run to $z=2.5$.

\section{Dark Galaxies}\label{sec:dark_galaxies}

Our previous work in Paper I was limited to $z \gtrsim 4$, but we have since extended the simulations to $z=2.5$. The left panel of Figure \ref{fig:fstar} shows a scatter plot of the stellar mass fraction ($f_\star = M_\star / \Mhalo$) against the total halo mass $\Mhalo$. Only halos in which the refinement reached the maximum refinement level of $\ell_{\rm max} = 7$ are analyzed, and there are 1010 such halos in our simulation at $z=2.5$. At high halo masses ($>10^{11} \Msun$) about half of the available gas has been converted to stars. Such a large star formation efficiency appears to be in conflict with observational constraints derived from abundance matching \citep[e.g.][]{moster_constraints_2010,behroozi_average_2012}, which imply a peak efficiency of a few percent for $10^{12} \Msun$ halos. This discrepancy is a well known problem with hydrodynamic galaxy formation simulations that include only a weak form of thermal supernova feedback \citep[see e.g.][]{katz_dissipational_1992}, and much contemporary work is concerned with solving this problem \citep{stinson_making_2013,hopkins_stellar_2012,agertz_towards_2012,guedes_forming_2011}, primarily by improving the treatment of feedback from young massive stars, supernovae, and black holes. Here we defer addressing this problem and instead wish to draw attention to the fact that a suppression of the gas-to-star conversion efficiency in low mass halos can be effected simply by the difficulty that low metallicity gas has in transitioning to a cold, molecular phase that is capable of forming stars.

In our simulation, the stellar mass fraction drops rapidly below $10^{10} \Msun$ and becomes highly stochastic, with $\fstar$ spanning about two orders of magnitude. The mean star formation efficiency drops from 0.046 to $1.6 \times 10^{-3}$ when averaged over all halos with mass $10^{10} - 10^{11} \Msun$ and $10^9 - 10^{10} \Msun$, respectively. A small number of halos (8 out of 1010) has not been able to form any stars, and we have marked these as upper limits at the lowest $\fstar$ that our simulation can resolve given our minimum stellar particle mass of $m_{\rm sp} = 10^4 \Msun$, of which $f_{\rm ej} = 0.25$ is ejected back into the interstellar medium as part of the supernova feedback prescription.

The middle panel of Figure \ref{fig:fstar} shows the total SFR vs. halo mass. Again the effect of our metallicity-dependent \HH-regulated SF prescription is clearly visible as a drop in SFR around $\Mhalo = 10^{10} \Msun$. At higher masses, the SFR falls on a narrow ``SFR main sequence'' corresponding to a roughly constant halo-mass-specific SFR of $\approx 4 \times 10^{-11} \, {\rm yr}^{-1}$. At lower halo masses, the SFR drops by almost two orders of magnitude, but remains non-zero owing to our 1 per cent \HH-floor. Slightly less than half of our halos (396 out of 1010) have zero SFR, and one tenth of these are plotted as upper limit arrows at the lowest SFR our simulation can resolve, $7.5 \times 10^{-4} \Msun$ yr$^{-1}$ (i.e. forming 1 star particle over the SF averaging time scale of 10 Myr).

Finally, the right panel of this figure shows the neutral hydrogen gas depletion time, defined as $\tdep = (M_{\rm HI} + M_{\rm H_2})/SFR$. On the ``SFR main sequence'' ($\Mhalo \gtrsim 10^{10} \Msun$) we find gas depletion times ranging from 0.4 to a few Gyr, consistent with the observational constraints from \citep{leroy_star_2008,bigiel_star_2008,genzel_study_2010}. The SF-suppressed galaxies make up a second population with depletion times in excess of 20 Gyr. Owing to their low stellar content and suppressed SFR, such galaxies would be extremely difficult to detect with continuum IR, optical, or UV observations, and hence we refer to them as ``dark galaxies''. In this work we use $\tdep > 20$ Gyr as a working definition of a dark galaxy. With this definition 789 out of our 1010 halos are dark galaxies. To first order, dark galaxies lie in halos that rely on the 1 per cent \HH-floor to form any stars, while halos that become metal enriched enough to enter the KMT regime ($\SigmaGas > \Sigma_{\rm crit}(Z)$) quickly form a siginficant stellar population and are luminous.

\begin{figure}[tp]
\includegraphics[width=0.93\columnwidth]{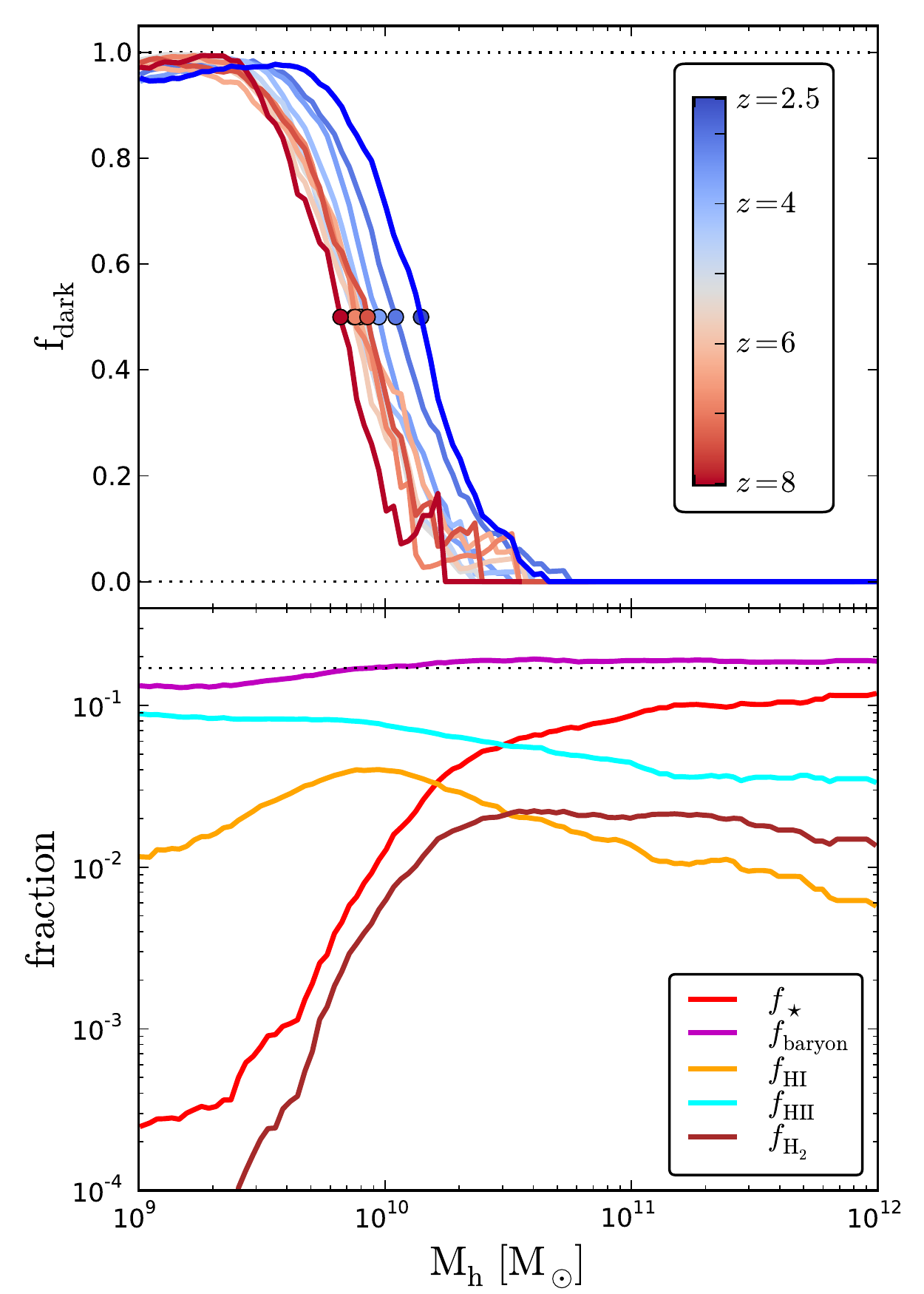}
\caption{Top panel: The fraction of dark galaxies \mbox{($\tdep > 20$ Gyr)} vs. total halo mass at different redshifts. The circles mark the halo mass at which $\fdark = 0.5$. Bottom panel: The average mass fraction of total baryons, neutral, ionized, and molecular hydrogen, and stars vs. total halo mass, at $z=2.5$. These lines were calculated using a sliding window of width 0.5 dex. The cosmic baryon fraction ($\Omega_b/\Omega_M$) is indicated by the dotted horizontal line. }
\label{fig:fdark+baryons}
\end{figure}

The top panel of Figure~\ref{fig:fdark+baryons} gives $\fdark$, the fraction of dark galaxies ($\tdep > 20$ Gyr) at a range of redshifts. At $z=2.5$, $\fdark$ rises from zero at $5 \times 10^{10} \Msun$ to near unity at $\Mhalo \lesssim 5 \times 10^9 \Msun$. The transition from mostly luminous to mostly dark halos shifts to lower halos masses at higher redshifts. A remarkable prediction of our work then is that the universe should be filled with a large number of dark galaxies of mass $\lesssim 10^{10} \Msun$ that were unable to convert their gas into stars simply because their metal content never rose to the level required for a conversion from atomic to molecular phase. Unlike in the more conventional feedback-based star formation suppression models, in our picture many of these dark galaxies would retain nearly the cosmic baryon budget in the form of neutral hydrogen, making them potentially observable in 21cm or Lyman alpha fluorescence. Indeed, \citet{cantalupo_detection_2012} may have already discovered examples of such dark galaxies in Ly$\alpha$ fluorescence illuminated by a $z \approx 2.4$ quasar.

The bottom panel of Figure \ref{fig:fdark+baryons} shows the mean baryonic contents of our halos as a function of total halo mass. As discussed above, the drop in $\fstar$ occurs around $10^{10} \Msun$, and it closely follows the transition between predominantly atomic gas in lower mass systems and mostly molecular gas in higher mass ones. At lower halo masses the effects of the metagalactic UV background are apparent, by lowering the neutral relative to ionized hydrogen abundance and a reduction in the total baryon fraction.

\begin{figure}[tp]
\includegraphics[width=\columnwidth]{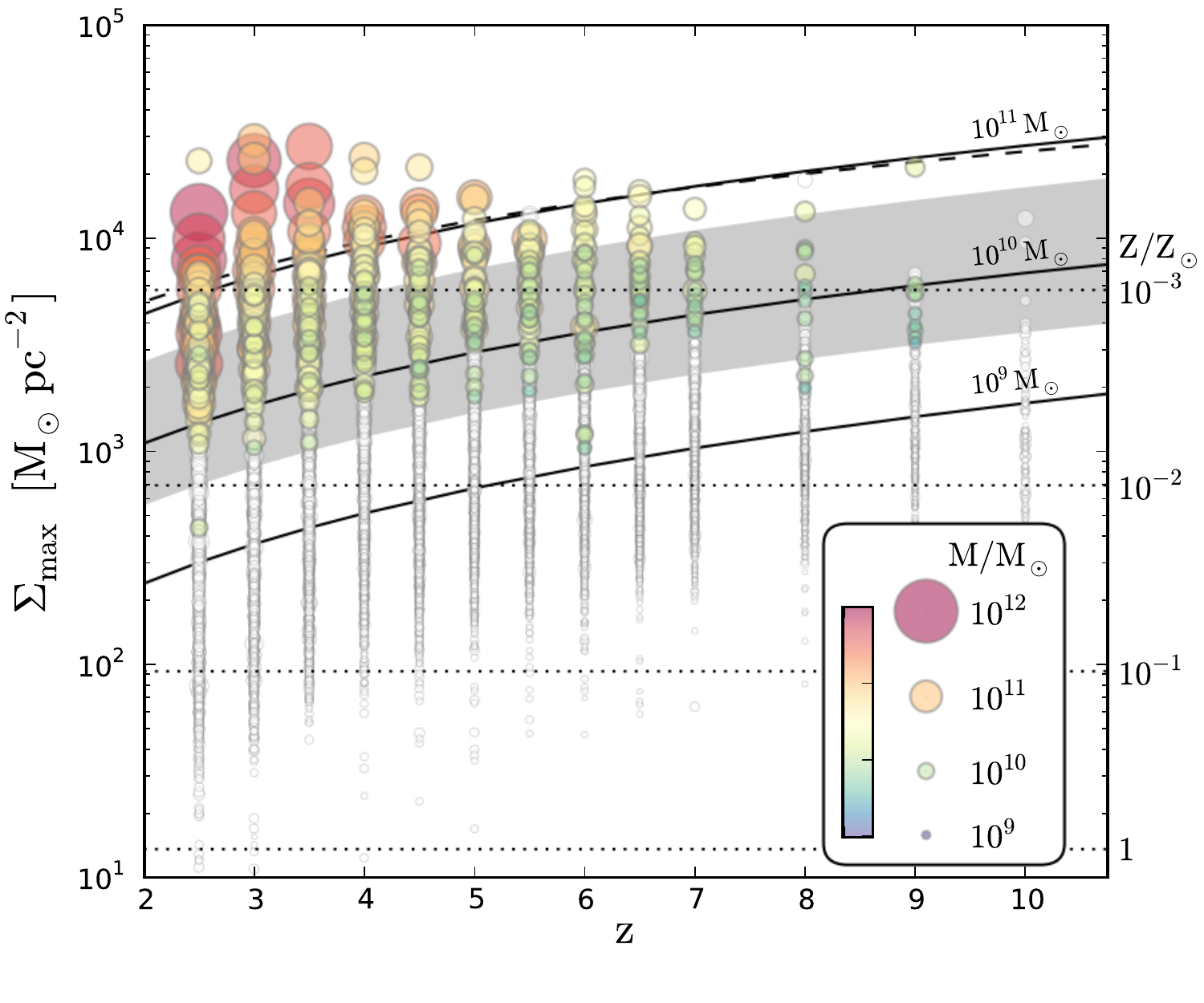}
\caption{The maximum column density, $\Sigma_{\rm max} = \rho_{\rm max} \times \Delta x_7 = \rho_{\rm max} \times 109 \, {\rm pc} \, (3.5/(1+z))$, reached in the simulated halos as a function of redshift. The size and color of the plotted symbol corresponds to the mass of the halo. Luminous galaxies are shown in color, dark galaxies in white. The horizontal dotted lines give the critical column density required for \HH\ formation, at metallicities (from top to bottom) $Z/\Zsun = 10^{-3}, 10^{-2}, 10^{-1}, \, {\rm and} \, 1$. The curved solid lines gives estimates of the maximum central surface density (averaged over $\Delta x_7$) calculated using the analytical model by \citet{mo_formation_1998} for halos of mass $10^{11}, 10^{10}, \, {\rm and} \, 10^{9} \Msun$ (see text for details). The solid lines are for a spin parameter $\lambda = 0.035$, and the gray shaded band delimits the region between $\lambda = 0.02$ and 0.05 for the $10^{10} \Msun$ case. The dashed line gives the analytical estimate for a fixed central averaging scale of 50 pc. This plot shows that (i) at a given redshift, more massive halos are more easily able to convert gas to the molecular phase and hence form stars, and (ii) at a given halo mass, it is easier for gas to become molecular at higher redshift, which results in an increase towards lower redshifts of the critical halo mass below which star formation is not possible in very low metallicity systems.}
\label{fig:Sigma_max}
\end{figure}

In our model, a halo's ability to become luminous depends on the maximum HI surface density its gas can reach. To demonstrate this, we show in Figure \ref{fig:Sigma_max} a plot of the maximum \HI\ density in each halo, for a range of redshifts. To get a simple estimate of the column density we multiply by the width of the most refined grid cell, $\Sigma_{\rm max} = \rho_{\rm max} \times \Delta x_7 = \rho_{\rm max} \times 109 \, {\rm pc} \, (3.5/(1+z))$. This quantity is plotted as circles whose size indicates the halo's mass and which are filled blue if the halo is luminous or white if it is a dark galaxy. $\Sigma_{\rm max}$ should be compared against the KMT critical column density to allow the atomic-to-molecular conversion to occur, which is metallicity dependent and plotted with dotted horizontal lines for metallicities $Z/\Zsun = 10^{-3}, 10^{-2}, 10^{-1}, \, {\rm and} \, 1$: $\Sigma_{\rm crit} = 5700, \, 690, \, 92.8, \, {\rm and} \, 13.6\, \Msun \, {\rm pc}^{-2}$. The lowest metallicity ($10^{-3} \Zsun$) corresponds to the metallicity floor that we impose at $z=10$, implying that halos must at some point exceed this threshold in order for a substantial number of stars to form. Subsequent star formation is much easier, since the metal enrichment from these stars rapidly lowers the required density. Halos that never exceed the $10^{-3} \Zsun$ column density threshold are dark galaxies. The figure illustrates two points, (i) that more massive halos are more easily able to convert atomic into molecular gas and hence support star formation and (ii) that the threshold halo mass for star formation increases towards lower redshifts (as also demonstrated in the top panel of Figure~\ref{fig:fdark+baryons}).

\section{Discussion}

\subsection{Analytical model}

To gain an understanding of the trends seen in Figure \ref{fig:Sigma_max}, we have calculated the expected central surface density using the analytical model of \citet{mo_formation_1998}. In this model the baryonic disk is completeley specified by its host halo's properties, and is modeled as an exponential disk embedded in the potential of an NFW dark matter halo\footnote{We neglect the treatment of adiabatic contraction, since observationally there does not seem to be much evidence for it in late type galaxies \citep[e.g.][]{dutton_dark_2011}.}. The disk mass is set to $m_d = 0.05$ times the total halo mass, and the disk scale radius is determined in an iterative manner from the assumption that the baryons contain a fraction $j_d = 0.05$ of the total angular momentum $J$ of the dark matter halo, which itself is related to the halo spin parameter $\lambda = J |E|^{1/2} G^{-1} M^{-5/2}$. Numerical simulations have shown that $\lambda$ is distributed approximately log-normally, with a mean of $\lambda \approx 0.035$ and a dispersion of $\sim 0.5$ \citep{knebe_correlation_2008}. Given this analytic disk model, we can estimate the central gas surface density on a length scale comparable to the most refined grid cells in our simulation. The solid lines in Figure \ref{fig:Sigma_max} give $\langle \Sigma\rangle_{\!R} = [\int_0^R 2 r \, \Sigma(r) \, dr]/R^2$ for $R = \Delta x_7$ in halos of mass $10^{11}, \, 10^{10}, \, {\rm and} \, 10^9 \Msun$ with a concentration $c=5$. The shaded band denotes the region spanned by $\lambda = 0.02$ and 0.05.

The analytical model agrees with our simulation results quite well, and explains why a halo of a fixed mass and metallicity has a harder time converting its gas to molecular phase at lower redshifts -- the disk surface density is set by the mean density of the halo, which decreases towards lower redshift. It follows that a halo with an initial metallicity equal to our metallicity floor of $Z/\Zsun = 10^{-3}$ can only become luminous if either it grows to a large enough mass that its disk surface density exceeds $\Sigma_{\rm crit} = 5700 \Msun \, {\rm pc}^{-2}$ or, alternatively, if it becomes externally enriched from a nearby star forming system. The star formation afforded by the low molecular gas content resulting from our 1 per cent \HH-floor by itself is generally too low to allow the halo to become luminous.

\subsection{External metal enrichment?}

The possibility of external enrichment deserves further consideration. As discussed above, our simulation has fairly weak stellar feedback, and as a result doesn't exhibit strong metal enriched galactic outflows. In order to estimate how many of our dark galaxies would have been externally enriched by more realistic galactic outflows from neighboring galaxies, we perform the following simple calculation: For every luminous halo we calculate its escape velocity ($v_{\rm esc} = (2 G \Mhalo/\Rvir)^{1/2}$), as well as the mean age of all its star particles, $\langle t_\star \rangle$. Assuming a galactic wind outflow with velocity $v_{\rm wind} = f_{\rm wind} \, v_{\rm esc}$, we determine the radius of its wind sphere of influence (SOI), $R_{\rm wind} = v_{\rm wind} \, \langle t_\star \rangle$. We estimate the mean metallicity of the wind by uniformly spreading the total metal mass contained in the luminous halo over the wind SOI. Next we determine for each dark galaxy how many SOI's it intersects, and sum up the metallicities of each of these to produce an estimate of the metallicity $Z_{\rm wind}$ that this dark galaxy would have been externally enriched to. We compare this metallicity to $Z_{\rm crit}\!(\Sigma_{\rm max})$, the KMT critical metallicity for atomic-to-molecular gas conversion given the maximum density reached in each of our dark galaxies. If $Z_{\rm wind} > Z_{\rm crit}$, then we judge this halo to have been externally enriched to the point where stars would have formed, and hence it would no longer be a dark galaxy. 

Using this procedure we find that for $f_{\rm wind} = 3$ only 7 out 789 dark galaxies would be externally enriched to above their critical metallicity. For $f_{\rm wind} = 1$ (5), this number becomes 44 (1). The fraction of externally enriched halos becomes smaller for larger $f_{\rm wind}$, because even though the enriched region extend farther and intercepts more dark halos, the wind SOI's metallicity becomes correspondingly smaller. It thus appears that the majority of our dark halos is isolated enough to remain unenriched, and hence dark, even in the presence of strong galactic outflows. Note that external enrichment will result in a strong spatial clustering of the small fraction of low mass halos that are able to become luminous in this way.

\begin{figure}[tp]
\includegraphics[width=\columnwidth]{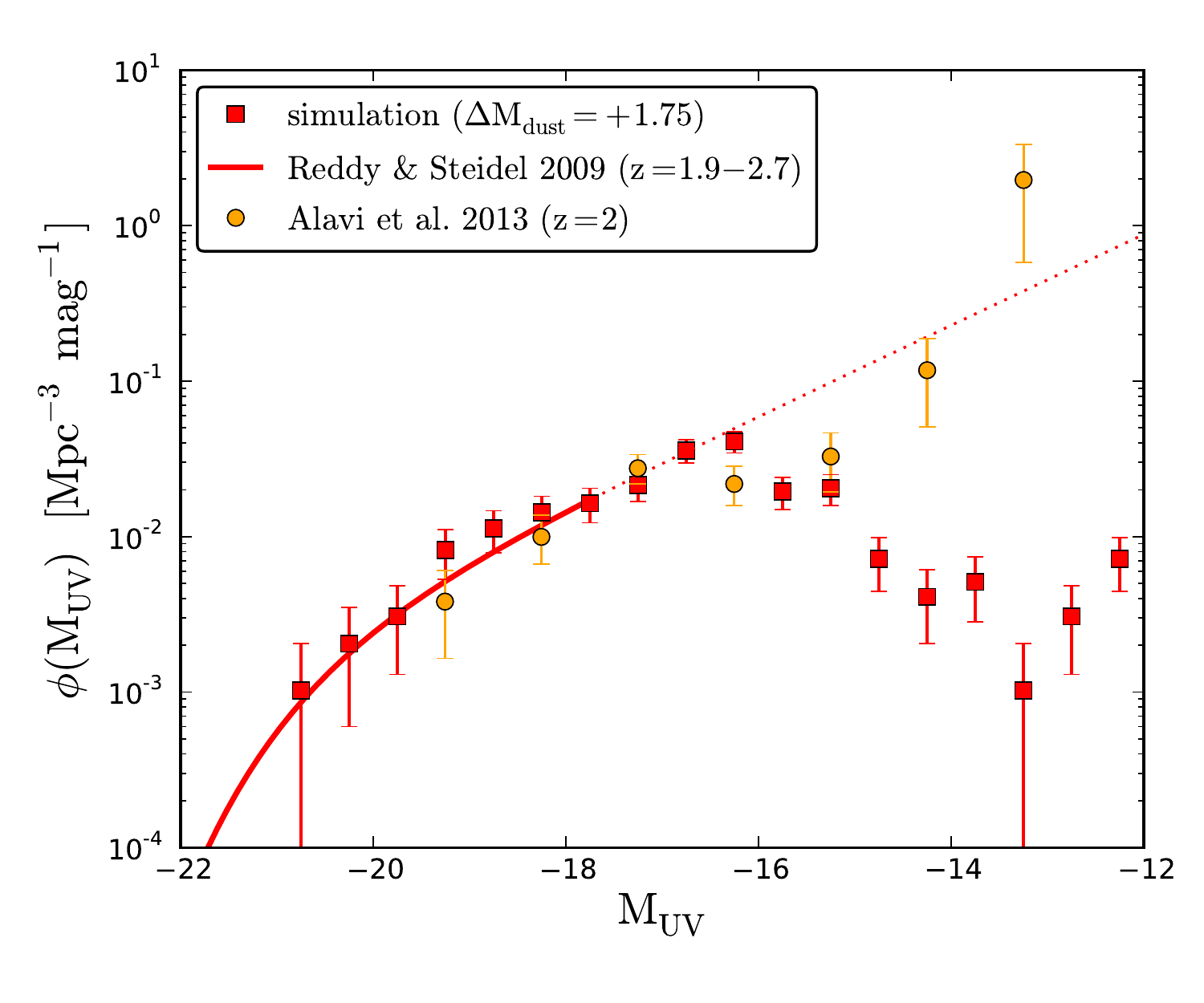}
\caption{The simulated UV luminosity function at \mbox{$z=2.5$} compared to the observational determinations from \citet{reddy_steep_2009} and \citet{alavi_ultra-faint_2013}. We calculate $\LUV$ from the SFR according to $\LUV = 1.25 \times 10^{28} \, ({\rm SFR}/\Msun\,{\rm yr}^{-1})$ erg s$^{-1}$ Hz$^{-1}$ (${\rm M_{UV}} = 51.63 - 2.5 \log_{10}(\LUV/$erg s$^{-1}$ Hz$^{-1})$), corresponding to a $Z = 0.2 \Zsun$ Chabrier IMF from 0.08 to 120 $\Msun$. The simulated galaxies' luminosities have been dimmed by a factor of 5 ($\Delta M = +1.75$) to account for the expected attenuation from dust, consistent with the observational constraints from \citet{reddy_goods-herschel_2012}. Note that the faintest two data points from \citet{alavi_ultra-faint_2013} consist of only 4 systems, and are likely subject to considerable systematic uncertainties arising from the lens mass modeling.}
\label{fig:LF}
\end{figure}

%%%and are likely affected by systematic uncertainties in the lens model magnification 

\begin{figure*}[htp]
\includegraphics[width=\textwidth]{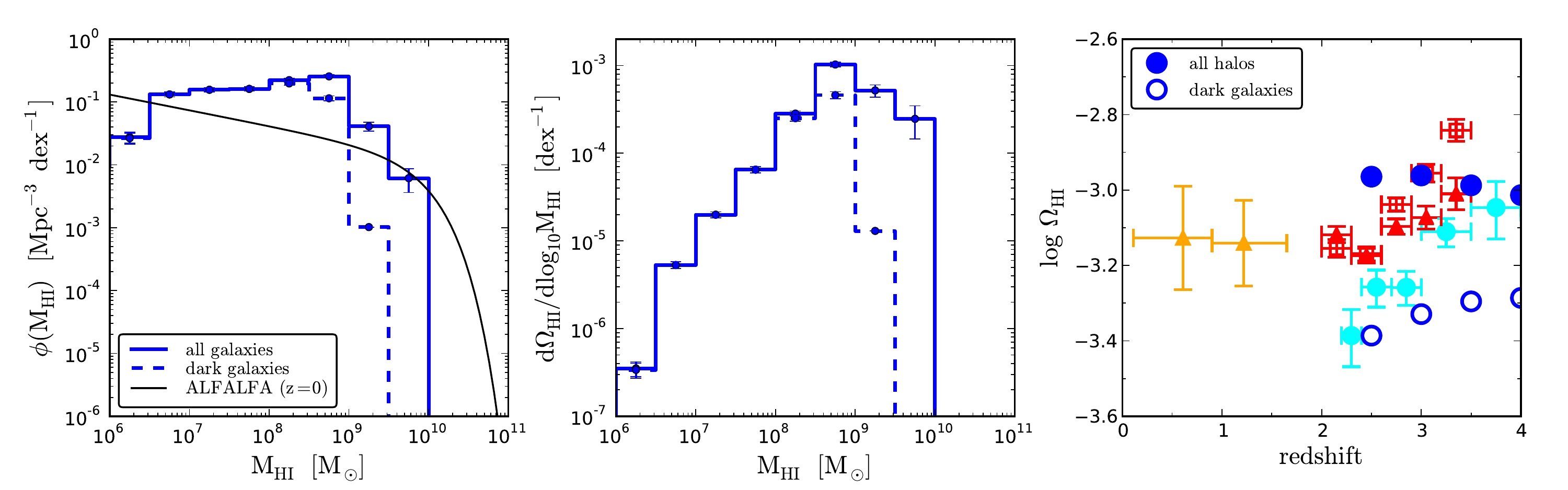}
\caption{Left: The simulated $\rm HI$ mass function at \mbox{$z=2.5$} compared to the $z \approx 0$ determination from the ALFALFA survey \citep{haynes_arecibo_2011}. The solid line is for all simulated halos, and the dashed for all dark galaxies. Middle: The differential contribution to $\OmegaHI$ per decade of $\MHI$ from all halos (solid) and dark galaxies (dashed). Right: $\OmegaHI$ vs. redshift. Comparison of observational constraints from \citet{rao_damped_2006} (orange triangles) at $z<2$ and \citet{prochaska_nonevolution_2009} (cyan circles) and \citet{noterdaeme_column_2012} (red symbols, the filled triangles are systematics-corrected) at $z > 2$ with $\OmegaHI$ determined in our simulation, summing over all halos (solid circles) and dark halos only (open circles).}
\label{fig:HI}
\end{figure*}

\subsection{$z=2.5$ UV luminosity function}

Next we confront our simulated galaxy population with the observationally determined rest-frame UV luminosity function (LF) of Lyman break galaxies (LBG) at $z=1.9 - 2.7$ from \citet{reddy_steep_2009} and the very recent photometric measurement at $z \approx 2$ by \citet{alavi_ultra-faint_2013}. For this purpose we calculate UV luminosities from the simulated SFR assuming a $Z = 0.2 \Zsun$ Chabrier initial mass function (IMF) from 0.08 to 120 $\Msun$ and a constant SFR for $\gtrsim 100$ Myr \citep{madau_star_1998}: $\LUV = 1.25 \times 10^{28}$ (SFR / $\Msun$ yr$^{-1}$) erg s$^{-1}$ Hz$^{-1}$. A Salpeter IMF would result in $\sim 1.7$ times lower luminosities. We obtain SFR from our simulated galaxies by summing the mass of all young star particles with ages less than $\tau_\star = 10$ Myr and dividing by this SF time scale, ${\rm SFR} = \sum_{{\rm age}\,<\,\tau_\star} m_\star/\tau_\star$. Star forming galaxies at $z \approx 2.5$ are known to be significantly reddened and extincted by internal dust obscuration. \citet{reddy_goods-herschel_2012} use a combination of \textit{Herschel}, \textit{Very Large Array}, and \textit{Spitzer} data of 146 UV-selected galaxies with spectroscopic redshifts in the GOODS-North field to derive a median dust correction factor of $5.2 \pm 0.6$ between the bolometric luminosity and the unobscured (i.e. observed) UV luminosity. Consistent with this measurement, we dim our SFR-derived UV luminosities by a factor of 5 ($\Delta M_{\rm dust} = +1.75$), and then calculate a restframe LF by binning our galaxies in bins of width 0.5 $\rm M_{UV}$ (Fig.~\ref{fig:LF}).

The agreement between the simulated and observed LF is remarkably good down to the spectroscopic observational limit of -18 mag from \citet{reddy_steep_2009}. Owing to our fairly small boxsize we don't have any simulated galaxies with dust-corrected UV luminosities brighter than -21 mag. The suppression of star formation due to the inability of low metallicity gas to become molecular in low mass halos results in a flattening and eventual cutoff in the faint end of the UV LF, which occurs 1-2 magnitudes below the current observational limit from spectroscopic studies. The agreement with the observational LF at $z \approx 2.5$ may be puzzling, given the unrealistically high $\fstar$ in our most massive halos. The only way for these two facts to be consistent is for the SFR to have been too high in the past. Indeed, Fig.~18 of Paper I indicates that the high redshift ($z \gtrsim 4$) LFs in our simulations exceed the observational constraints.

Very recent photometric work by \citet{alavi_ultra-faint_2013} has extended the $z \sim 2$ LF to several magnitudes fainter systems, by taking advantage of the large magnifications provided by the strongly lensing galaxy cluster Abell 1689. Their work indicates a steeply rising faint end of the LF down to $M_{\rm UV} \approx -13$, in apparent conflict with the predictions of our \HH-regulated SF simulation. We caution, however, that the faintest two data points ($M_{\rm UV} > -15$) of the \citet{alavi_ultra-faint_2013} LF (the only ones disagreeing with our predictions) are comprised of only 4 systems, and are likely subject to considerable systematic uncertainties, due to the difficulty of determining accurate lensing magnifications, which affects both the inferred intrinsic magnitudes and the effective volume probed by the survey. Furthermore, the disagreement with the faintest two data points would be greatly reduced, if we applied a luminosity-dependent dust attenuation law \citep[see e.g.][]{reddy_steep_2009} and dimmed $M_{\rm UV} > -15$ galaxies by less than a factor of 5. For these reasons, we don't consider the current observational LF constraints to rule out the existence of a large number of dark galaxies at $z \approx 2.5$. Our simulations, however, predict that the faint-end of the UV LF should not continue to rise much beyond $M_{\rm UV} \approx -16$. The \citet{alavi_ultra-faint_2013} study highlights the potential of future deeper multicolor imaging surveys to constrain the \HH-regulated SF scenario.

\subsection{HI abundance and mass function}\label{sec:HI}

In Figure \ref{fig:HI} we consider the abundance and mass distribution of neutral hydrogen in our simulation. The left panel shows the \HI\ mass function at $z=2.5$. We find halos with \HI\ masses as low as $50 \Msun$ and as high as $4.6 \times 10^9 \Msun$. From $\MHI=10^7$ to $10^9 \Msun$ the \HI\ mass function is remarkably flat (in contrast to the DM mass function which is steeply rising towards low masses), and exhibits a sharp cutoff at higher \HI\ masses. This cutoff is a result of the limited box size of our simulation, which prohibits us from resolving halos more massive than $\approx 10^{12} \Msun$ at $z=2.5$. The flat shape of the mass function at lower masses is due to the ionizing and heating effects of the metagalactic UV background, which causes a declining neutral gas fraction towards lower mass halos (see Fig.~\ref{fig:fdark+baryons}), thus compensating the higher abundance of such halos. Dark galaxies make up the vast majority of all $\MHI < 10^9 \Msun$ systems. Comparing to the observational measurement of the \HI\ mass function in the local universe from the ALFALFA survey \citep{haynes_arecibo_2011}, we see an excess of about 1 order of magnitude at $\MHI$ between $10^7$ and a few $\times 10^9 \Msun$ in our simulation. Furthermore, since our simulation overproduces the stellar content of the brightest galaxies (see \S~\ref{sec:dark_galaxies}), those same galaxies have values of $\Mstar / \MHI$ that are too high compared to the relation observed in present day galaxies \citep{huang_gas_2012}.

Note that we are here comparing $z \approx 0$ observational measurements with simulation results from an epoch $\sim 11$ Gyr earlier, so this is not really a fair comparison. Nevertheless the excess of low $\MHI$ systems is somewhat troubling, since most physical processes that are commonly proposed to lower the \HI\ content in low mass halos \citep[e.g. radiative feedback from young stars or supernova driven galactic outflows,][]{altay_through_2011,dave_neutral_2013} rely on a substantial SFR, which is not present in our dark galaxies. It is possible that over the following 11 Gyr the low residual star formation in our dark galaxies (stemming from the \HH-floor), or external enrichment from nearby luminous halos, will raise the metal content of these systems to the point where the molecular transition and hence more efficient star formation becomes possible. Additionally, interactions with the cosmic web could ram pressure strip gas from these dark galaxies \citep{benitez-llambay_dwarf_2013} and reduce their \HI\ content.

In the central panel of Figure \ref{fig:HI} we show the differential contribution to $\OmegaHI$ per decade of $\MHI$ (i.e. $\MHI \, \phi(\MHI) / \rho_{\rm crit}$) from all halos and from dark halos only. In our simulation, the dominant contribution to $\OmegaHI$ comes from systems with $\MHI$ between $3 \times 10^8$ and $10^9 \Msun$, and luminous halos contribute about 1.5 times as much to the total \HI\ budget than dark galaxies. The shape of our distribution closely matches that of the observational distribution from the ALFALFA survey \citep[see Fig.~9 of][]{martin_arecibo_2010}, which, however, is peaked at about half a dex higher $\MHI$. Again, we attribute this shift to the comparatively small boxsize of our simulation.

In the right panel we show a comparison of our simulation results to observational measurements of $\OmegaHI$ from quasar absorption line studies from \citet{rao_damped_2006} at $z<2$ and \citet{prochaska_nonevolution_2009} and \citet{noterdaeme_column_2012} at $z>2$. The total \HI\ content of our simulation is in reasonable agreement with the data. At $z=2.5$ it is about $0.2-0.4$ dex (factor of $1.5 - 2.5$) too high, and this is likely a consequence of the weak supernova feedback in our simulation \citep[see also][]{altay_through_2011,erkal_origin_2012,dave_neutral_2013}. We re-emphasize that the dark galaxies in our simulation only contribute sub-dominantly to the total \HI\ content of the simulation (see the open circles in the figure), and the absence of radiative feedback or supernova-driven outflows in these systems is not a problem for the \HI\ budget.

\begin{figure}[tp]
\includegraphics[width=\columnwidth]{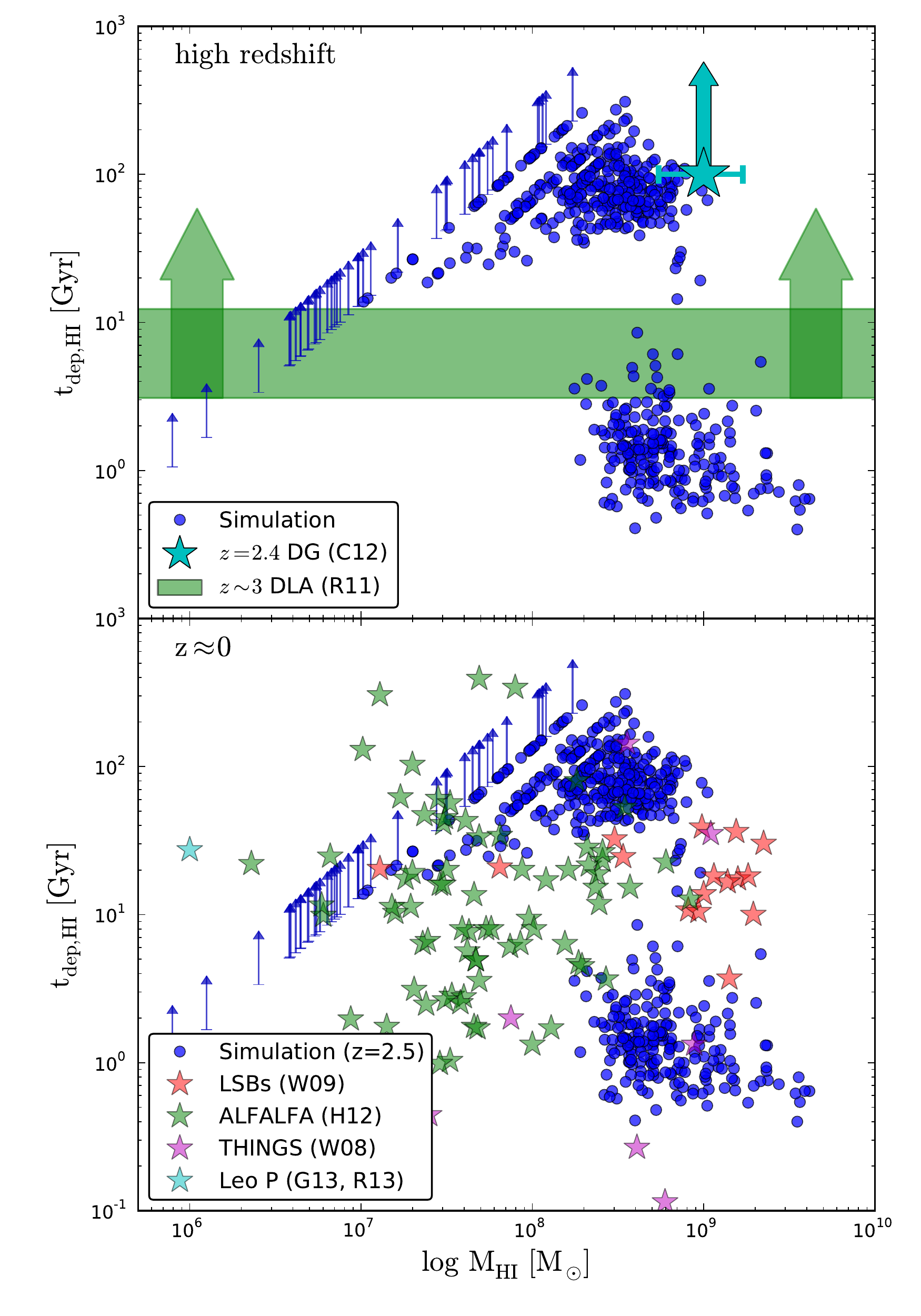}
\caption{Comparison of the relation between \HI\ depletion time $\tdepHI$ and $\MHI$ with observational data. Top panel: High redshift data. Simulation points are from the $z=2.5$ output. As in Figure \ref{fig:fstar} we mark halos with $\rm SFR=0$ as lower limits at the depletion time corresponding to the minimum resolvable SFR in our simulation. The cyan star is the lower limit from the stack of dark galaxies in \citet[C12]{cantalupo_detection_2012} assuming a SFR $< 10^{-2} \Msun \, {\rm yr}^{-1}$ and their $\MHI$ estimates (width of the errorbar). The green shaded band indicates the range of lower limits to the depletion times reported by \citet[R11]{rafelski_star_2011} in $z \sim 3$ DLA's without a central SF component. Bottom panel: Low redshift observational data. Simulation data is the same as in the top panel. The data points (star symbols) are for low surface brightness galaxies \citep[red,][W09]{wyder_star_2009}, ALFALFA dwarf galaxies \citep[green,][H12]{huang_gas_2012}, the THINGS survey \citep[magenta,][W08]{walter_things:_2008}, and for the newly discovered Leo P dwarf galaxy \citep[cyan,][R13, G13]{rhode_alfalfa_2013, giovanelli_alfalfa_2013}.
All data, both local and high-redshift, have been standardized to the star formation rate-luminosity calibrations recommended by \citet{kennicutt_star_2012}.
}
\label{fig:tdep_vs_MHI}
\end{figure}

\subsection{\HI\ depletion time vs. $\MHI$}

Finally, in Figure \ref{fig:tdep_vs_MHI} we show a comparison of the relation between \HI\ depletion time and \HI\ mass in our simulation at $z=2.5$ to observational data both from high redshift ($z \approx 2.5-3$, top panel) and in the local universe (bottom panel). In the top panel we include the ``dark galaxy'' systems reported by \citet{cantalupo_detection_2012}, which were detected as Ly$\alpha$ fluorescence in a $z \approx 2.4$ quasar field. Assuming a negligible contribution to the Ly$\alpha$ luminosity from internal SF or cooling, these authors derive a total hydrogen mass of $\sim 10^9 \Msun$ (assumed to be mostly ionized). Without a nearby quasar to ionize it, most of this hydrogen would be neutral and so we equate it with $\MHI$. The range of $\MHI$ derived from the Ly$\alpha$ luminosities is indicated by the horizontal error bar. Since no continuum emission was measured in these systems, they report only an upper limit of SFR$ < 0.01 \Msun \, {\rm yr}^{-1}$ derived from a stacking analysis, and a corresponding lower limit on $\tdepHI$. This lower limit is consistent with the long depletion times of our simulated dark galaxies, but systems with $\MHI \gtrsim 10^9 \Msun$ in our simulation tend to lie in $>10^{10} \Msun$ halos with more active star formation and shorter depletion times. Note, however, that the \HI\ masses derived from the unresolved Ly$\alpha$ sources are quite uncertain, since they depend on modeling the unknown spatial distribution of the fluorescing gas. 

We also plot in this panel a band corresponding to \HI\ depletion times determined in $z \approx 3$ damped Lyman alpha (DLA) systems without central bulges of star formation from \citet{rafelski_star_2011}.  Since no length scales are known for these DLAs, we cannot obtain $\MHI$ estimates and instead show a horizontal band. The vertical extent of the band corresponds to the range of smoothing parameters employed by \citet{rafelski_star_2011}, which result in different lower limits on $\tdepHI$. The band lies above the cloud of simulation points corresponding to actively star forming halos with short $\tdepHI$, and is consistent with the much longer $\tdepHI$ of our dark galaxies.

In the lower panel of this plot we compare to measurements in the local universe. Again we caution that this is not really a fair comparison, since a lot of evolution in both SFR and $\MHI$ may occur in the $\sim 11$ Gyr between $z=2.5$ and today. Nevertheless, it is interesting to see how our $z=2.5$ dark galaxies compare to observational constraints in $z=0$ dwarf galaxies. We plot data points from low surface brightness galaxies \citep{wyder_star_2009}, ALFALFA dwarf galaxies \citep{huang_gas_2012}, the THINGS survey \citep{walter_things:_2008}, and for the newly discovered Leo P dwarf galaxy \citep{rhode_alfalfa_2013}. We see that a small fraction of the local dwarfs overlap with the cloud of high $\tdepHI$ simulation points, indicating that some counterparts of dark galaxies may have already been observed in the local universe. Most of the $z \approx 0$ dwarfs, however, fall in between the bimodal distribution of simulation points, with intermediate $\tdepHI$ of $\sim 10$ Gyr, where there are few systems in our simulation. The sharpness of the division between low and high $\tdepHI$ systems is likely to be a consequence of the unique value of our \HH\ floor and our adoption of a single $Z_{\rm floor}$; the latter issue has been emphasized by \citet{tassis_ultra-faint_2012} and \citet{krumholz_metallicity-dependent_2012}. In reality there is probably some scatter in both the minimum \HH\ content in low column density gas and the metal content provided by population III star formation, and incorporating this scatter in our simulation would fill in some of the intermediate $\tdepHI$ region. Furthermore, any reduction in the \HI\ content of our dark galaxy halos, which we already know to be necessary from our comparison with the $z=0$ ALFALFA \HI\ mass function (see Section~\ref{sec:HI}), would move the high $\tdepHI$ points down and to the left, bringing them into better agreement with the local dwarf data. A pure downward shift of these points could be effected by a slight increase in the SFR at a fixed \HI\ content, which may occur through gradual internal or external metal enrichment.

\section{Summary and Conclusions}

We have extended the work of \citet{kuhlen_dwarf_2012}, by evolving simulations with a metallicity-dependent \HH-regulated SF prescription for an additional one billion years to $z=2.5$. We have improved the simulations by imposing a one per cent floor in the \HH\ fraction of cold metal poor gas, where the KMT prescription predicts zero molecular gas content. While the previous simulations exhibited a sharp cutoff in the stellar content below $10^{10} \Msun$, our new simulations results in a more gradual turnover, with star formation becoming effectively stochastic at halo masses below $10^{10} \Msun$. We have shown that this stochasticity reflects the halo formation time, with more massive (at a fixed time) and earlier collapsing (at a fixed mass) halos more easily exceeding the metallicity-dependent surface density threshold ($\SigmaGas \gtrsim 5700 \Msun \, {\rm pc}^{-2}\, 
(Z/10^{-3}\Zsun)^{\!-0.88}$) for molecular gas formation. If the resulting $\sim 2$ dex scatter in the stellar mass content of $\lesssim 10^{10} \Msun$ halos persists to $z=0$, it may explain the puzzling dearth of bright dwarf satellite galaxies in the Local Group. In this picture, massive halos that previously have been deemed ``too big to fail'' \citep{boylan-kolchin_milky_2012}, would instead just be halos lying in the tails of the $\fstar$ distribution.

An interesting implication of our results is that the $z \approx 2.5$ universe may be filled with a large population of gas rich halos with very low stellar content and close to zero ongoing star formation, which we term ``dark galaxies''. We define dark galaxies to be systems with a neutral gas depletion time of $\tdep = M_{\rm HI + H_2} / {\rm SFR}$ greater than 20 Gyr. In our simulation we find that the fraction of dark galaxies rises from zero at $\Mhalo > 5 \times 10^{10} \Msun$ to near unity at $\Mhalo < 1 \times 10^{10} \Msun$. In total 78 per cent of resolved halos (789 out of 1010) are classified as dark galaxies.

Remarkably, the existence of a substantial population of dark galaxies at $z=2.5$ appears to be fully compatible with existing observational constraints. In particular our simulated galaxies matches the luminosity function determined by \citet{reddy_steep_2009}, once their luminosities as dimmed by a factor of 5 \citep[consistent with measurements by][]{reddy_goods-herschel_2012} to account for dust dimming. Our simulation also reproduces to within a factor of $\sim 2$ the cosmological mass density of neutral gas at $z \gtrsim 2.5$.

We conclude by highlighting some areas where our simulations are challenged by observational data, and which demand additional work. The biggest concern is the stellar mass content in halos sufficiently massive ($\Mhalo > 10^{11} \Msun$) to avoid being constrained by the molecular gas transition. The lack of efficient feedback processes in our simulation allows almost half of all gas to be converted into stars, resulting in a stellar mass fraction that is 5 - 10 times higher than observational constraints based on abundance matching \citep{moster_constraints_2010,behroozi_average_2012}. A second, possibly related, concern is that, compared to the \HI\ mass function measured by the ALFALFA survey in the local universe \citep{martin_arecibo_2010,haynes_arecibo_2011}, we overproduce by almost one order of magnitude the abundance of $\MHI < 10^9 \Msun$ systems. The majority of this excess is contributed by dark galaxies, for which more effective stellar or supernova feedback would not likely help much, given their low SFR. It remains to be seen whether internal or external processes in the subsequent 11 Gyr between $z=2.5$ to $z=0$ can reduce the neutral gas content in low mass halos, and if so, whether these processes maintain the stochastic nature of the stellar mass content in the present universe.

\acknowledgments
We thank Naveen Reddy for assistance with the dust attenuation of high redshift star forming galaxies, and Sebastiano Cantalupo, Phil Hopkins, and Eliot Quataert for stimulating discussions. Support for this work was provided by the NSF through grants OIA-1124453 and AST-0955300, by NASA through ATP grants NNX12AF87G, NNX13AB84G, and a Chandra telescope grant, and by the Alfred P.~Sloan Foundation. We acknowledge computational support from the NASA Advanced Supercomputing Division, on whose \textit{Pleiades} supercomputer this simulation was carried out.

\bibliographystyle{apj}
\bibliography{DarkGalaxies}

\end{document}